\documentclass{aa}  

\usepackage{graphicx}
\usepackage{txfonts}
\usepackage{xcolor}
\usepackage[flushleft]{threeparttable}
\usepackage{changes}
\usepackage{natbib}
\begin{document}

   \title{On the TeV Halo Fraction in gamma-ray bright Pulsar Wind Nebulae}

   \subtitle{}

   \author{G. Giacinti\inst{1}, A.M.W. Mitchell\inst{2}, R. L\'{o}pez-Coto\inst{3}, V. Joshi\inst{4}, R.D. Parsons\inst{1}, J.A. Hinton\inst{1}}
   \authorrunning{G. Giacinti et. al.}

   \institute{Max-Planck-Institut f\"{u}r Kernphysik, P.O. Box 103980, D 69029 Heidelberg, Germany \label{MPIK} \and  Physik Institut, Universität Zürich, Winterthurerstrasse 190, CH-8057 Zürich, Switzerland \label{UZH} \and  Universit\`{a} di Padova and Istituto Nazionale di Fisica Nucleare, I-35131, Padova, Italy \label{Padova} \and Friedrich-Alexander-Universit\"{a}t Erlangen-N\"{u}rnberg, Erlangen Centre for Astroparticle Physics, Erwin-Rommel-Str. 1, D 91058 Erlangen, Germany \label{ECAP}}

 
  \abstract
   {

   The discovery of extended TeV emission around the Geminga and PSR\,B0656+14 pulsars, with properties consistent with free particle propagation in the interstellar medium (ISM), has sparked considerable discussion on the possible presence of such halos in other systems. Here we make an assessment of the current TeV source population associated with energetic pulsars, in terms of size and estimated energy density. Based on two alternative estimators we conclude that a large majority of the known TeV sources have emission originating in the zone energetically and dynamically dominated by the pulsar (i.e. the \emph{pulsar wind nebula}), rather than from a halo of particles diffusing in to the ISM. 
   Furthermore, whilst the number of established halos will surely increase in the future, we find that it is unlikely that 
   such halos contribute significantly to the total TeV $\gamma$-ray luminosity from electrons accelerated in PWN.
   }

   \maketitle
%

\section{Introduction}
\label{sec:intro}

The size of the zone within which a pulsar's influence is dominant, i.e. the pulsar wind nebula (PWN), is dictated primarily by the injected energy and the external pressure. Within this zone relativistic particle propagation may be dominated by advection, rather than diffusion. Particle acceleration to $>$TeV energies is thought to occur at or near the pulsar wind termination shock, located at $\sim 0.1$\,pc from the pulsar, but relativistic particles are present, and radiating, on (potentially) much larger scales. The usual formalism applied to study the evolution of a PWN is magneto-hydrodynamics (MHD) and in this description the escape of particles cannot easily be included, but it is generally accepted that for young pulsars, such as the Crab, the confinement within the PWN is very effective. At some point in the evolution of the PWN, mixing of the low density PWN material with the ISM or the reverse shock of an associated supernova remnant (SNR) is likely, the PWN boundary becomes less clear cut and escape becomes possible. Escaping particles encounter the conditions of the ISM (or perhaps modified conditions associated with the earlier SNR, cf. \cite{evoli_linden_morlino}) and their transport becomes dominated by diffusion, potentially forming a detectable halo of escaped particles, and/or reaching the Earth to contribute to the measured local electron and positron fluxes. The recent detection of extended TeV emission around the two very nearby low power pulsars Geminga and  PSR\,B0656+14 \citep{geminga_hawc_paper,2hwc_catalog} has led to an extensive discussion of \lq\lq TeV halos\rq\rq\/ and their contribution to the general population of TeV emitting PWN \citep{invisible_pulsars_linden}.

We consider a reasonable definition of an electron halo (referred to as `halo' hereafter) to be the presence of an over-density of relativistic electrons around a source/acceleration-site, in a zone in which the source itself does not dominate the dynamics or composition of the interstellar medium. This implies that diffusion is always expected to dominate in halos unless sources exist in regions where there are very large scale flows, such as a galactic wind or outflow. We note that halos may not always be symmetric. Electrons follow magnetic field lines in the interstellar turbulence around the source, and, in cases where the bulk of emitting electrons is located at distances from the source that are smaller than the coherence length of the turbulence, the halo is expected to appear asymmetric and filamentary~\citep{mag_trb_geminga}.

The formation of a halo is possible around any source class from which cosmic-ray electrons may escape before cooling. For GeV-emitting electrons this has long been assumed to be the case for SNR, for example. For the $>$10 TeV electrons producing TeV $\gamma$-rays, radiative lifetimes are much shorter, $\sim 10^{4} (B/10\,\mu\rm{G})^{-2} (E_{e}/10\,\rm{TeV})^{-1}$ years, and it is less clear if escape before cooling is possible for SNR, or for other galactic sources. 

\begin{figure*}
\centering
\includegraphics[width=0.85\linewidth]{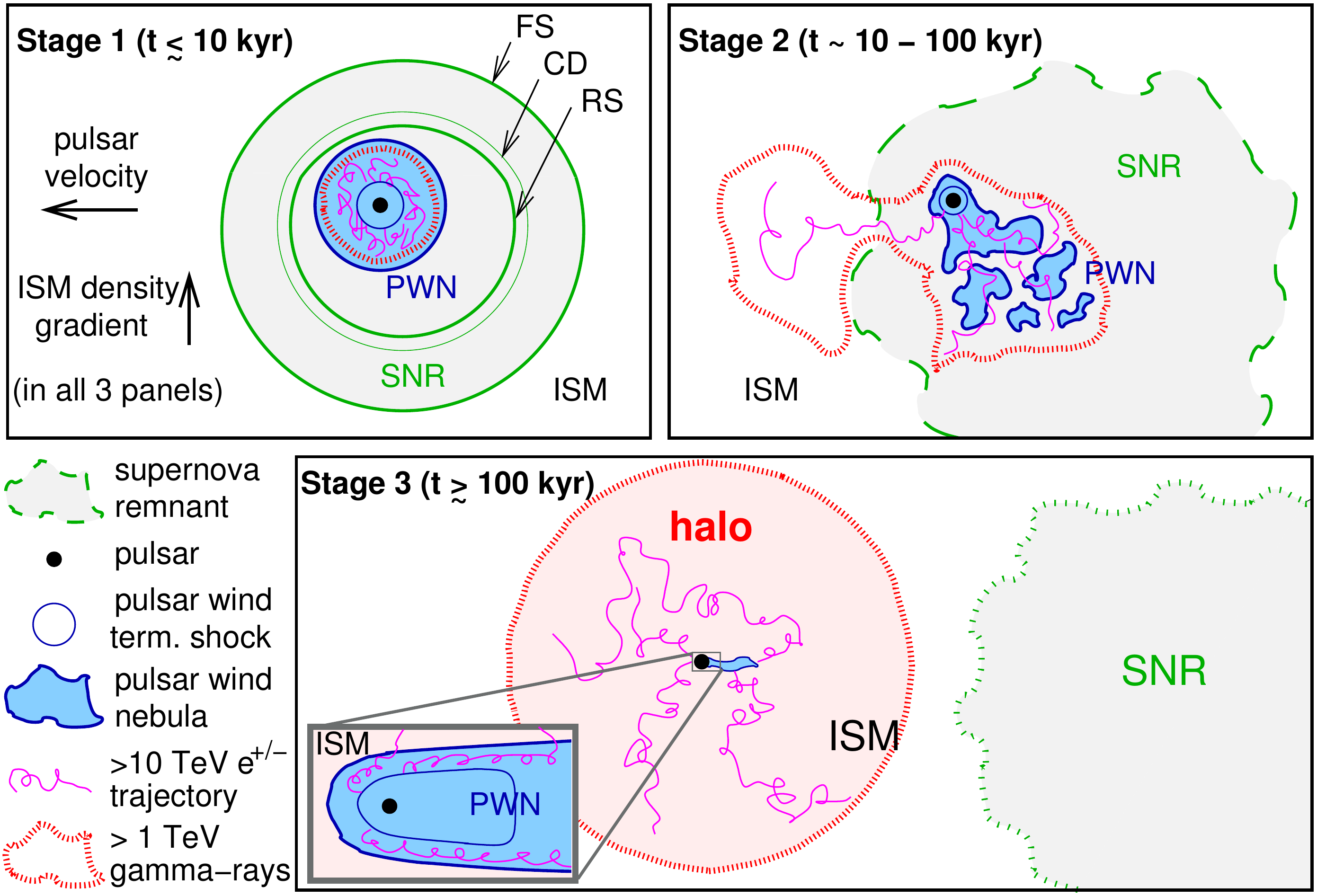}
\caption{Sketch of the main evolutionary stages of a PWN. The upper left panel shows early times, $t\lesssim 10$\,kyr (\lq\lq stage 1\rq\rq\/), when the PWN is contained inside the SNR and before the reverse shock (RS) interacts with it. The SNR forward shock (FS) and contact discontinuity (CD) are plotted with green lines. The electrons that are responsible for the TeV gamma-ray emission of the nebula are thought to be confined within the nebula at this stage. The upper right panel shows intermediate times, $t\sim 10 - 100$\,kyr (\lq\lq stage 2\rq\rq\/), after the PWN is disrupted by the reverse shock, but before the pulsar escapes its SNR. At this stage, TeV gamma-ray emitting electrons start to escape from the PWN, into the SNR and possibly into the ISM. The lower panel depicts the system at late times, $t\gtrsim 100$\,kyr (\lq\lq stage 3\rq\rq\/), when the pulsar has escaped from its ---now fading--- parent SNR. At this stage, high-energy electrons escape into the surrounding ISM, and may, only then, form a halo. See the text in Section~\ref{sec:PWN_evolution} for more details. The key is in the lower left corner. In all three panels, the ISM density gradient is upwards, and the pulsar \lq\lq kick\rq\rq\/ velocity towards the left.}
\label{fig:pwn_halo_sketch}
\end{figure*}

Pulsar wind nebulae are the only class of sources for which the escape of TeV-emitting electrons is now firmly established~\citep{geminga_hawc_paper}, after long being discussed as a potential source of the locally measured cosmic ray electrons \citep[see e.g.][]{aharonian_atoyan}. There are also indications of missing (apparently escaped) high energy electrons in the case of the Vela system~\citep{velax_escape}, however alternative explanations such as rapid cooling due to an enhanced magnetic field are possible \citep{VelaXERN}.
Apparently escaped relativistic electrons are detected in X-rays around the Guitar nebula \citep[e.g.][]{Johnson_2010_Guitar_Nebula} and the Lighthouse nebula \citep[e.g.][]{Pavan_2014}. The most obvious way to differentiate between escaped and confined radiating particles is to estimate the PWN boundary using multi-wavelength data. This process is often problematic due to the effects of instrumental sensitivity, non-uniform magnetic fields and particle cooling. For example, it is very often the case that the zone in which imaging of a PWN is possible in the X-ray domain, with instruments such as Chandra or XMM-Newton, is very much smaller than the physical size of the PWN as determined in other wavelengths.  Within the X-ray domain, the physical PWN size is also often energy dependent, which is interpreted as a signature of the rapid cooling of the highest energy electrons producing the keV synchrotron emission. 
Indeed, the typical cooling time of electrons emitting photons with characteristic energy $h\nu_{\rm c}$ is $\sim 10^{3}\,{\rm yr}\,(B/10\,\mu{\rm G})^{-3/2}(h\nu_{\rm c}/5\,{\rm keV})^{-1/2}$. In the radio domain, the cooling effect is unimportant, but surface brightness sensitivity is usually sufficient only for young and compact sources.

Here we consider various estimates of the expected size of the nebulae around pulsars that have been associated to TeV emission, comparing these estimates to the measured sources sizes. We also assess the fraction of the power that is present in sources with and without halos and hence their contribution to the total gamma-ray emission of all pulsars within star-forming systems.
%

\section{Pulsar Wind Nebula Evolution}
\label{sec:PWN_evolution}
According to the above definition, halos may exist only around PWN whose electrons and positrons have started to escape into the surrounding, unperturbed ISM. It is therefore instructive to recall briefly the main stages of the evolution of a PWN.
The environment of pulsars changes dramatically over time, firstly as contained within an evolving supernova remnant (SNR), and finally within the general ISM when the \lq\lq kick\rq\rq\/ velocity received by the pulsar at birth moves it beyond the decelerated shell of the host SNR. There is considerable literature associated with PWN evolution, including several reviews, see in particular \citet{gaensler_slane_review}. In general, however, the existing work focuses on X-ray and radio, rather than TeV emission, and/or exclusively on the early to middle ages ($\ll 100$\,kyr) of PWN evolution. Here we consider briefly the physical properties of the region from which TeV emission originates during the lifetime of a pulsar.

Figure~\ref{fig:pwn_halo_sketch} illustrates three stages in the evolution of a TeV-emitting PWN.
We depict in chronological order: first, the system at early times $t \lesssim 10$\,kyr after the supernova in the upper left panel, then intermediate times $t \sim 10 - 100$\,kyr in the upper right panel, and, finally, late times $t \gtrsim 100$\,kyr in the lower panel. Hereafter, we refer to these three stages as \lq\lq stage 1\rq\rq\/, \lq\lq stage 2\rq\rq\/, and \lq\lq stage 3\rq\rq\/, respectively. In all three panels of this sketch, the \lq\lq kick\rq\rq\/ velocity that is initially imparted to the pulsar during the supernova explosion is assumed to point towards the left, and the ISM density gradient in which the SNR evolves to point \lq\lq upwards\rq\rq\/. The areas shaded in grey correspond to the SNR, and the surrounding ---solid, dashed or dotted--- green lines denote the location of its forward shock. The black dots show the location of the pulsar, the PWN is shaded in blue, and the pulsar wind termination shock is represented with the thin solid blue line inside the PWN. The inset in the lower panel corresponds to an enlargement of the innermost regions of the PWN in stage 3. The high-energy electrons and, or, positrons that are responsible for the TeV emission from PWN are usually thought to be accelerated at the termination shock. We note that electrons and positrons are often assumed to be accelerated in equal quantities, but this may not be the case. \cite{GG_Kirk_ApJ_2018} found that an individual pulsar may favour either positrons or electrons at the highest energies. Hereafter we use the word \lq\lq electrons\rq\rq\/ indiscriminately when referring to electrons and/or positrons, unless stated otherwise. In each panel, we sketch with thin magenta lines a few illustrative trajectories for some of these $\gtrsim 10$\,TeV electrons. The dotted red lines delineate the typical extent of the $\gtrsim 1$\,TeV gamma-ray emission resulting from these electrons (see Figure~\ref{fig:pwn_halo_sketch}).


\subsection{Stage 1: $<$ 10\,kyr}

At early times (stage 1), the pulsar is still relatively close to its birthplace. The relativistic electron-positron wind that surrounds it, and which is decelerated to non-relativistic speeds at the termination shock, inflates a nebula inside the parent SNR. The ejecta from the supernova expands at supersonic speeds in the surrounding ISM, creating a forward shock (labelled \lq\lq FS\rq\rq\/ ). Inside the volume delimited by the contact discontinuity (\lq\lq CD\rq\rq\/) between the ejecta and the shocked ISM, a reverse shock (\lq\lq RS\rq\rq\/) appears as the expansion slows down. At these early times $t \lesssim 10$\,kyr, the reverse shock has not yet reached the PWN. The high-energy electrons accelerated at the PWN are thought to remain confined inside at this stage, and the TeV gamma-rays emitted by these electrons should therefore come from within the nebula. 
Therefore, TeV emission from PWNe in stage 1 cannot be associated to halos. For the sake of simplicity, we draw the nebula as a spherically symmetric system, but we note that the reality may be more complex, see e.g. the Crab Nebula where prominent filaments from the Rayleigh-Taylor instability are present. Eventually, the reverse shock returns inwards, towards the centre of the explosion. As is depicted in Fig.~~\ref{fig:pwn_halo_sketch}, the SNR expands more slowly in the direction where the ISM density is higher, and the reverse shock may then reach the PWN earlier from some directions than in others. At the end of stage 1 and the beginning of stage 2, the reverse shock crushes and disrupts the PWN; against which the PWN forward shock rebounds, with the system experiencing several reverberations between the shock and the PWN \citep{Blondinpwnsnr}.

\subsection{Stage 2: $10-100$\,kyr}

At intermediate times $t \sim 10-100$\,kyr (stage~2, see Figure~\ref{fig:pwn_halo_sketch}), the morphology of a PWN-SNR system is often highly irregular. It depends both on the properties of the material in the surrounding ISM, and on the direction and velocity of the pulsar. The nebula is disrupted and the pulsar can be strongly off-centre with respect to the PWN. In our sketch, we draw the typical geometry one would expect for a system evolving in a smooth region of the ISM with a steady density gradient perpendicular to the pulsar velocity. (See, for example, Figures~12 and 13 in~\cite{Slane_Vela_X} for hydrodynamical simulations of Vela~X, as well as Figure~6 in~\cite{g327_temim_xray} and Figure~8 in~\cite{Kolb_2017} for simulations of G\,327.1$-$1.1). At this stage, high-energy electrons start to escape from the PWN, and propagate in to the surrounding SNR, with further escape into the surrounding ISM becoming possible. A study of electron escape from Vela~X is presented in~\cite{velax_escape}. In Figure~\ref{fig:pwn_halo_sketch}, we show that the 
extent of the resulting TeV emission may now be greater than that of the nebula. This still does not constitute a true TeV halo, in that the bulk of the TeV emission originates from electrons that do not propagate in the \lq\lq unperturbed\rq\rq\/ ISM. We note that the sketch in Figure~1 of~\cite{TeV_halos_everywhere} corresponds to a standard PWN-SNR system in its stage~2, albeit in the specific case of a homogeneous ISM density and no pulsar motion.

\subsection{Stage 3: $>$ 100\,kyr}

At late times, typically $t \gtrsim 100$\,kyr (stage 3, see Figure~\ref{fig:pwn_halo_sketch}), the pulsar has finally escaped from its parent SNR, due to its kick velocity. At this stage, the SNR is expanding very slowly and fading away. The pulsar propagates in the ISM, and forms a bow-shock PWN, with a tail of shocked pulsar wind trailing behind (blue region in Figure\,\ref{fig:pwn_halo_sketch}). The inset shows an enlargement of the region around the head of the nebula. Observations of X-ray \lq\lq filaments\rq\rq\/ around some bow-shock PWNe demonstrate that high-energy electrons can escape into the surrounding ISM. See e.g.~\cite{Johnson_2010_Guitar_Nebula} and \cite{Hui_2012_Guitar_Nebula} for the Guitar Nebula, and \cite{Pavan_2014,Pavan_2016} for the Lighthouse Nebula. Theoretical studies also suggest that some of the high-energy electrons inside a bow-shock PWN can escape into the ISM, even in regions that are not far from the head of the nebula: see e.g.~\cite{Bucciantini_2018} and \cite{Barkov_2019}. These particles should then diffuse in the surrounding turbulent interstellar magnetic fields and emit TeV gamma-rays in a volume that is substantially larger than that of the PWN. 

Only at this stage, and under the condition that escaped relativistic electrons do not dominate the energy density of the ISM, should the TeV emission be considered as a \lq\lq TeV halo\rq\rq\/. In general, only old PWNe with ages $\gtrsim 100$\,kyr (or, at least, $\gtrsim$ a few tens of kyr) may then be surrounded by TeV halos. The recent HAWC detection of halos at TeV energies around the two stage~3 PWNe of Geminga pulsar and PSR B0656+14 \citep{geminga_hawc_paper} is consistent with this picture.

\begin{table*}
\begin{center}
\begin{threeparttable}
\caption{Properties of selected well-known PWN systems in different evolutionary stages, ordered according to the system age.}
\begin{tabular}{l c c c c c c c c}
\hline \hline
System & Crab & MSH\,15-52 & G21.5-0.9 & G0.9+0.1 & Vela X & G327.1-1.1 & J1825-137  & Geminga\\
\hline    
Age (kyr)\tnote{a} & 0.94 & 1.56 & 4.85 & 5.31 & 11.3 & 18 & 21.4 & 342 \\
PSR\tnote{b} & {\small B0531+21} & {\small B1509-58} & {\small J1833-1034} & {\small J1747–2809} & {\small B0833-45} & \tnote{c} & {\small B1823-13} & {\small J0633+1746} \\ %
$\log(\dot E)$ (erg/s)& $38.65$ & $37.23$ & $37.53$ & $37.63$  & $36.84$ & $36.49$ & $36.45$  & $34.51$ \\
Distance (kpc)& 2 & 4.4 & 4.1 & 8.5 & 0.28 & 9 & 3.93  & 0.25\\
R$_{\rm SNR}$ (pc)& ?\tnote{d} & 38.4 & 2.98 & 19.8 & 19.5 & 22 & 120  & ?\\
R$_{\rm PWN}$ (pc)\tnote{e} & 2.8 & 19.2 & 0.8 & 2.5 & 12.2 & 10.5 & ?  & 0.01\\
$v \times t$ (pc)\tnote{f}& 0.27 & 0.45 & 1.4 & 1.5 & 3.3 & 5.2 & 6.2 & 100\\
R$_{\rm TeV}$ (pc)\tnote{g} & < 3 & 11 & < 4 & < 7 & 2.9 & 3 & 50  & 16.2\\
R$_{\rm X-ray}$ (pc)& 0.24 & 10.2 & 0.8 & 4.9 & 3.08 & 13 & 9.1  & 0.15\\
Stage\tnote{h} & 1 & 1 & 1b & 1b & 2 & 2 & 2b  & 3 \\
Refs.\tnote{i} &  I & II & III & IV & V & VI & VII & VIII \\
\hline
\end{tabular}
\begin{tablenotes}
\small
\item[a] The pulsar characteristic age is used for the age of the system, except where historical values are known. 
\item[b] Associated pulsar (PSR). Pulsar properties are taken from \cite{Manchester05}.
\item[c] Putative pulsar candidate identified, without pulsed emission detected \cite{g327_temim_pulsar}
\item[d] Unknown quantities are marked by \lq\lq ?\rq\rq\/ 
\item[e] R$_{\rm PWN}$ is the size of the PWN in radio (as opposed to the radio SNR shell). 
\item[f] $v \times t$ is the pulsar kick velocity multiplied by the age of the system, where a value of 300 km/s is adopted for the velocity, corresponding to the average of known values \citep{psr_avg_kick_vel}. 
\item[g] R$_{\rm TeV}$ is the one sigma radius taken from \cite{pwn_pop_paper} for sources within the H.E.S.S. Galactic Plane Survey (HGPS), unless a reference is provided. 
\item[h] Stage of system evolution is assigned loosely based on age, to correspond to Fig. \ref{fig:pwn_halo_sketch}
\item[i] References: {\bf I:} \cite{crab_snr_frail,crab_pwn_kargaltsev_review} {\bf II:} \cite{msh1552_caswell_snr,msh1552_duplessis,msh1552_trussoni_xray,msh1552_mineo_xray} {\bf III:} \cite{g21_matheson_xray,g21_safiharb_xray} {\bf IV:} \cite{Green_snr_cat,green_cat2,g09_dubner_radio,g09_porquet_xray} {\bf V:} \cite{velax_duncan_snr,velax_dwarakanath_radio,velax_tibaldo} {\bf VI:} \cite{G327_ma_radio,g327_temim_xray} {\bf VII:} \cite{j1825_stupar_snr,j1825_duvidovich_radio,j1825_pavlov_xray,j1825_uchiyama_xray} {\bf VIII:} \cite{geminga_pellizzoni_radio,geminga_hawc_paper,geminga_posselt_xray,geminga_caraveo_xray}
\end{tablenotes}
\label{tab:halos_table}
\end{threeparttable}
\end{center}
\end{table*}

\subsection{Properties of known PWNe}

Table~\ref{tab:halos_table} provides a summary of the properties of several well-studied PWN systems in different evolutionary stages, roughly corresponding to the three stages illustrated in Figure~\ref{fig:pwn_halo_sketch}. Where the evolutionary stage is denoted as \lq\lq 1b\rq\rq\/ or \lq\lq 2b\rq\rq\/ in Table~\ref{tab:halos_table}, the system may be considered as currently between stages. This assignment is roughly based on the pulsar characteristic age, however this may be a poor estimate of the true system age; in particular, the morphology of HESS\,J1825-137 suggests transitory behaviour with evolutionary models favouring an older age for the system \citep{hessj1825_2019,Vanettenromani}.

To explore the relative sizes of the different components in a PWN-SNR system and their evolution, Table\,\ref{tab:halos_table} compiles multi-wavelength measurements of the emission extent. The size of the SNR, R$_{\rm SNR}$, is given by the radius of a shell, where detected - often from radio information. The size of the central PWN, R$_{\rm PWN}$, is obtained from the extent of radio emission located immediately around the pulsar. The X-ray size, R$_{\rm X-ray}$, provided for comparison is a measure of the central PWN size, which traces to the youngest energetic particles in the system.
For the synchrotron emission seen in the radio to X-ray ray the observed sizes relate to the magnetic field distribution, as well as the particle distribution. In contrast the TeV size, R$_{\rm TeV}$, depends only on the particle distribution, at least in the usual case of close to uniform radiation density for inverse Compton scattering.

At early evolutionary times (stage 1) both X-ray and TeV emission fill the available region for a similar overall size as the PWN remains confined by a shock front.
Towards later times (stages 2-3) the discrepancy between X-ray and TeV size becomes more pronounced, as IC scattering continues to produce significant TeV emission from older, lower energy particle populations long after the synchrotron X-ray emission has dropped below detectable levels due to cooling.
Typically, R$_{\rm TeV}$ will therefore be somewhat larger than R$_{\rm X-ray}$ for older systems, with $\mathrm{R}_{\rm TeV}/\mathrm{R}_{\rm X-ray} \approx 5$ for HESS\,J1825-137 and $\approx 100$ for Geminga, whilst for the younger system MSH\,15-52, $\mathrm{R}_{\rm TeV}/\mathrm{R}_{\rm X-ray} \approx 1$.
We note, however, that here $R_{\rm TeV}$ is given by the one sigma radius of the emission from \cite{pwn_pop_paper}, whereas $R_{\rm X-ray}$ is given by the total visible extent of X-ray emission. 

Lastly, as a further scale comparison, the pulsar kick velocity $v$ and age of the system $t$ are used to obtain an estimated displacement of the pulsar from its birth place.
In Table\,\ref{tab:halos_table} a value of 300\,km/s is adopted, corresponding to the average of known pulsar kick velocities \citep{psr_avg_kick_vel}. 

In comparing the emission extent for specific PWN-SNR systems across multiple wavelengths, differences in projection effects arising from variation in the extent along the line of sight have been neglected and are assumed to be small with respect to the distance to the system.

\section{Energy Density Estimates}
\label{energy_density_estimate}

Using two different methods, we now estimate the energy densities $\varepsilon_{\rm e}$ in relativistic electrons and positrons around pulsars with established associated TeV emission; see  Table~\ref{tab:population}. Our selection of TeV sources comprises those identified as PWNe by the HESS collaboration in their study of the PWN population \citep{pwn_pop_paper}, expanded to include  Geminga and PSR\,B0656+14 \cite{geminga_hawc_paper}.
Comparing these energy densities with the typical energy density of the ISM, $\varepsilon_{\rm ISM}$, allows one to determine in a first approximation whether the electrons that are responsible for the VHE $\gamma$-ray emission occupy the relatively unperturbed ISM ($\varepsilon_{\rm e} \ll \varepsilon_{\rm ISM}$), or if they are still contained in a region energetically and dynamically dominated by the pulsar ($\varepsilon_{\rm e} \gtrsim \varepsilon_{\rm ISM}$). %

\subsection{Estimator using pulsar properties}
\label{sec:method1}

Measured properties of the associated pulsar can be used to estimate the total power injected into accelerated electrons. The energy density is estimated simply as $\varepsilon_{\rm e} = E_{\rm inj}/V$, where $V=4\pi R^{3}/3$ and $R$ is the 68\% containment radius for the TeV emission. For Geminga and PSR\,B0656+14 we calculate, according to the morphology reported in \cite{geminga_hawc_paper}, the radius at which 68\% of the total gamma-ray emission is contained. For the remaining sources, the sizes reported in Tables 1 and 3 of \cite{pwn_pop_paper} are scaled to correspond to 68\% containment and upper limits are kept as reported. Two cases are considered: A: $E_{\rm inj} = \dot{E}_0 \tau_0$, where $\dot{E}_0$ is the initial spin down power and $\tau_0$ is the characteristic spin-down timescale; and B: $E_{\rm inj}=\dot{E}\tau_{\rm c}$, where $\tau_{\rm c}$ is the characteristic age of the pulsar and $\dot{E}$ the present spin-down power. 

Case A represents the case of injection of essentially the full rotational energy of the pulsar soon after birth. We adopt values for the initial characteristic spin-down timescale corresponding to an extreme value of the birth period $P_{0}=10$ ms, in order to provide an upper limit to the energy content.
This assumption leads to inferred energy densities larger than 10\,eV/cm$^3$ for all the sources considered here. Whilst for older sources, where cooling losses cannot be ignored, this is clearly an overestimate, it is worth noting that a low birth period can easily result in very high energy densities for systems of age $<$ a few 10s of kyrs.

\begin{figure*}
\centering
\includegraphics[width=\linewidth]{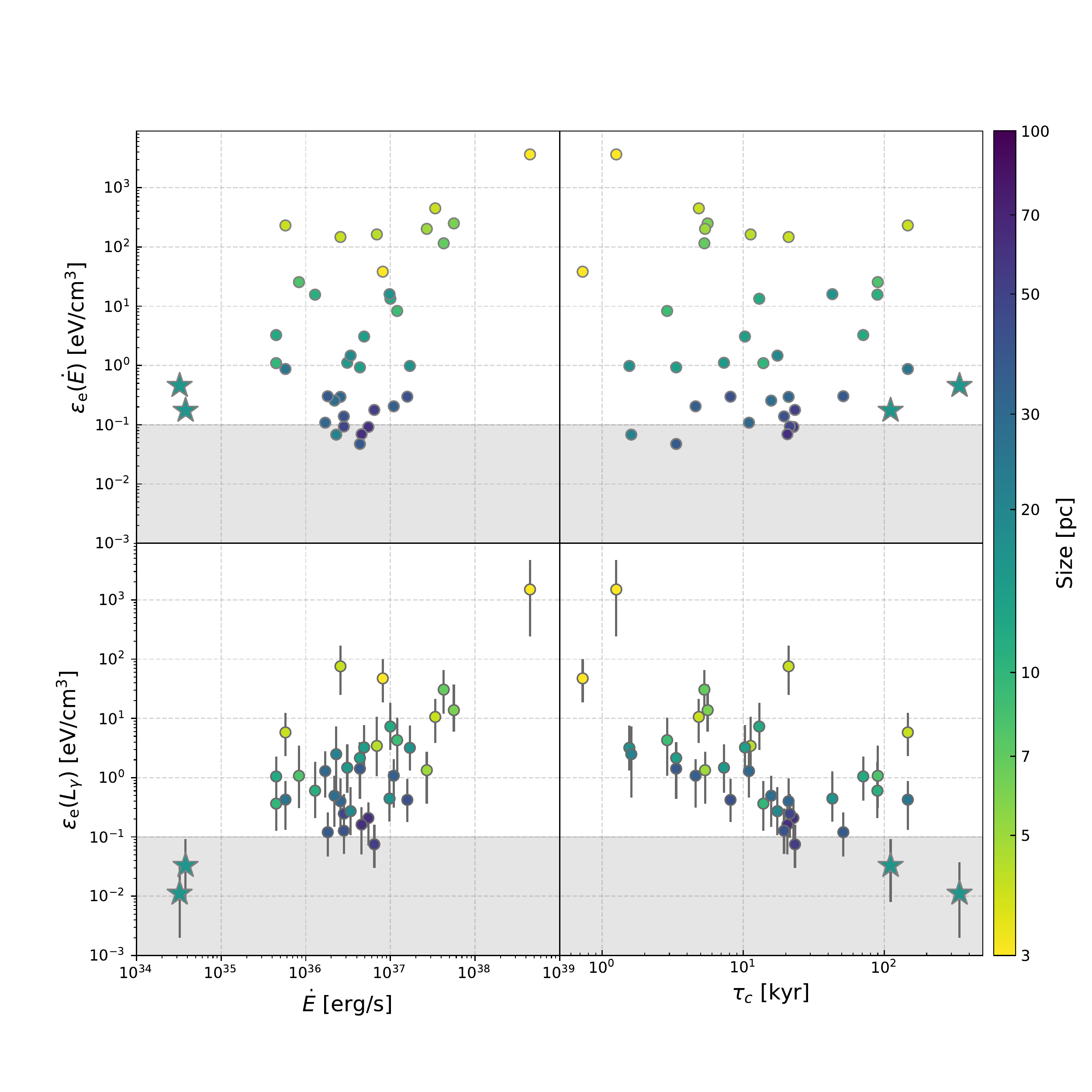}
\caption{Energy density within TeV sources associated with pulsars as a function of pulsar $\dot{E}$ and $\tau_c$ under different assumptions. The top panels represent the energy density calculated as $\varepsilon_{\rm{e}} = \dot{E} \tau_c/V$, described in section \ref{sec:method1}. The bottom panels represent the energy density calculated using the electron spectrum derived from TeV $\gamma$-ray measurements, described in section \ref{sec:method2}. The shaded regions correspond to an energy density lower than that of the ISM under any reasonable assumption. Systems for which measured properties are taken from the HGPS are indicated by circles, whereas stars indicate the systems where HAWC data is used (see also Table \ref{tab:population}). }
\label{fig:edensities}
\end{figure*}

In Case B, we adopt the current spin down power of the pulsar as a measure of power input, again neglecting  energy losses, hence: $E_{\rm inj} = \dot{E}\tau_{\rm c}$. Results with this estimator are shown in the two upper panels of Figure~\ref{fig:edensities} and in the second column from the right in Table \ref{tab:population}. The estimates in this case are typically a factor of $10-100$ lower than those of Case A. This case is pessimistic in the sense that it ignores the early evolution of the pulsar where significant energy injection may have taken place at higher $\dot{E}$, but again optimistic in ignoring losses and energy input to other channels (e.g. expansion of the nebula, magnetic fields). We consider Case B to be a better guide in general, particularly for evolved sources as losses are unlikely to be negligible for TeV gamma-ray emitting particles in the early life of the PWN when magnetic fields are high. 

Clearly more sophisticated treatments are possible, but all rely on models of the evolution of pulsars and their nebulae, which are not currently well constrained by the available data. Instead we turn to existing measurements at TeV energies to establish empirically the current day electron population in (and/or around) these PWN.

\begin{table*}
\begin{center}
\caption{ A compilation of TeV $\gamma$-ray sources with pulsar associations (i.e. PWN or associated Halos), together with estimated electron energy densities within the emission region (see notes below and text for details). 
}
\begin{tabular}{l l c c c c c c c c c}
\hline \hline
Name & ATNF Name & $\log_{10}\dot E$ & $\tau_{\rm c}$ & Distance & Size & $\log_{10}E_{\rm Total}$ & $\log_{10}\varepsilon_{\rm e}(\dot E)$ & $\log_{10}\varepsilon_{\rm e}(L_\gamma)$ \\
& & $\left(\rm erg/s\right)$  & (kyr) & (kpc) &  (pc) & $\left(\rm erg \right)$ & (eV/cm$^3$) & (eV/cm$^3$)\\
\hline

Geminga & J0633+1746 & 34.51 & 342 & 0.25$^{[1]}$ & 16.2 & 47.5 & $-0.3$ & $-2.0_{-0.8}^{+0.5}$ \\
B0656+14 & B0656+14  & 34.58 & 111 & 0.29$^{[2]}$ & 16.2 & 47.1 & $-0.8$ & $-1.5_{-0.6}^{+0.4}$ \\
CTA\,1 & J0007+7303 & 35.65 & 13.9 & 1.4 & 10.0 & 47.3 & 0.0 & $-0.4_{-0.5}^{+0.4}$ \\
J1858+020 & J1857+0143 & 35.65 & 71.0 & 5.8 & 11.9 & 48.0 & 0.5 & $0.0_{-0.4}^{+0.3}$ \\
J1834-087 & B1830-08 (2) & 35.76 & 147 & 4.5 & 25.7 & 48.4 & $-0.1$ & $-0.4_{-0.5}^{+0.3}$ \\
J1832-085 & B1830-08 (1) & 35.76 & 147 & 4.5 & 4.0 & 48.4 & 2.4 & $0.8_{-0.4}^{+0.3}$ \\
J1026-582 & J1028-5819 & 35.92 & 90.0 & 2.3 & 8.0 & 48.4 & 1.4 & $0.0_{-0.5}^{+0.5}$ \\
J1718-385 & J1718-3825 & 36.11 & 89.5 & 3.6 & 10.9 & 48.6 & 1.2 & $-0.2_{-0.5}^{+0.5}$ \\
J1303-631 & J1301-6305 & 36.23 & 11.0 & 6.7 & 31.1 & 47.8 & $-1.0$ & $0.1_{-0.4}^{+0.3}$ \\
J1809-193 (2) & J1809-1917 & 36.26 & 51.3 & 3.5 & 37.7 & 48.5 & $-0.5$ & $-0.9_{-0.4}^{+0.3}$ \\
J1804-216 & B1800-21 & 36.34 & 15.8 & 4.4 & 28.7 & 48.0 & $-0.6$ & $-0.3_{-0.5}^{+0.3}$ \\
J1119-614 & J1119-6127 & 36.36 & 1.61 & 8.4 & 21.1 & 47.1 & $-1.2$ & $0.4_{-0.7}^{+0.5}$ \\
J1018-589A & J1016-5857 (1) & 36.41 & 21.0 & 8.0 & 4.0 & 48.2 & 2.2 & $1.9_{-0.5}^{+0.4}$ \\
J1018-589B & J1016-5857 (2) & 36.41 & 21.0 & 8.0 & 31.7 & 48.2 & $-0.5$ & $-0.4_{-0.5}^{+0.4}$ \\
J1825-137 & B1823-13 & 36.45 & 21.4 & 3.9 & 48.3 & 48.3 & $-1.0$ & $-0.6_{-0.4}^{+0.3}$ \\
J1908+063 & J1907+0602 & 36.45 & 19.5 & 3.2 & 41.1 & 48.2 & $-0.9$ & $-0.9_{-0.4}^{+0.4}$ \\
J1356-645 & J1357-6429 & 36.49 & 7.31 & 2.5 & 15.2 & 47.9 & 0.0 & $0.2_{-0.4}^{+0.4}$ \\
J1708-443 & B1706-44 & 36.53 & 17.5 & 2.6 & 19.2 & 48.3 & 0.2 & $-0.6_{-0.4}^{+0.4}$ \\
J1641-462 & J1640-4631 (2) & 36.64 & 3.35 & 12.8 & 14.0 & 47.7 & 0.0 & $0.3_{-0.5}^{+0.3}$ \\
J1640-465 & J1640-4631 (1) & 36.64 & 3.35 & 12.8 & 37.7 & 47.7 & $-1.3$ & $0.2_{-0.5}^{+0.3}$ \\
J1857+026 & J1856+0245 & 36.66 & 20.6 & 9.0 & 61.9 & 48.5 & $-1.2$ & $-0.8_{-0.5}^{+0.3}$ \\
J1418-609 & J1418-6058 & 36.69 & 10.3 & 5.0$^{[3]}$ & 14.2 & 48.2 & 0.5 & $0.5_{-0.4}^{+0.4}$ \\
J1837-069 & J1838-0655 & 36.74 & 22.7 & 6.6 & 61.9 & 48.6 & $-1.0$ & $-0.7_{-0.5}^{+0.3}$ \\
J1809-193 (1) & J1811-1925 & 36.81 & 23.3 & 5.0 & 52.8 & 48.7 & $-0.8$ & $-1.1_{-0.4}^{+0.3}$ \\
J0835-455 & B0833-45 & 36.84 & 11.3 & 0.3 & 4.4 & 48.4 & 2.2 & $0.5_{-0.5}^{+0.5}$ \\
J1846-029 & J1846-0258 & 36.91 & 0.73 & 5.8 & 3.0 & 47.3 & 1.6 & $1.7_{-0.4}^{+0.3}$ \\
J1849-000 & J1849-0001 & 36.99 & 42.9 & 7.0$^{[4]}$ & 16.6 & 49.1 & 1.2 & $-0.4_{-0.4}^{+0.5}$ \\
J1420-607 & J1420-6048 & 37.00 & 13.0 & 5.6 & 11.9 & 48.6 & 1.1 & $0.9_{-0.4}^{+0.4}$ \\
J1023-575 & J1023-5746 & 37.04 & 4.60 & 8.0$^{[5]}$ & 35.0 & 48.2 & $-0.7$ & $0.0_{-0.5}^{+0.3}$ \\
J1930+188 & J1930+1852 & 37.08 & 2.89 & 7.0 & 9.0 & 48.0 & 0.9 & $0.6_{-0.6}^{+0.4}$ \\
J1616-508 & J1617-5055 & 37.20 & 8.13 & 6.8 & 42.3 & 48.6 & $-0.5$ & $-0.4_{-0.4}^{+0.4}$ \\
J1514-591 & B1509-58 & 37.23 & 1.56 & 4.4 & 16.8 & 47.9 & 0.0 & $0.5_{-0.4}^{+0.4}$ \\
3C58 & J0205+6449 & 37.43 & 5.37 & 2.0$^{[6]}$ & 5.0 & 48.7 & $2.3$ & $0.1_{-0.6}^{+0.3}$ \\
J1833-105 & J1833-1034 & 37.53 & 4.85 & 4.1 & 4.0 & 48.7 & 2.7 & $1.0_{-0.4}^{+0.3}$ \\
G0.9+0.1 & J1747-2809 & 37.63 & 5.31 & 13.3 & 7.0 & 48.9 & 2.1 & $1.5_{-0.4}^{+0.3}$ \\
J1813-178 & J1813-1749 & 37.75 & 5.60 & 4.7 & 6.0 & 49.0 & 2.4 & $1.1_{-0.4}^{+0.4}$ \\
Crab Nebula & B0531+21 & 38.65 & 1.26 & 2.0 & 3.0 & 49.2 & 3.6 & $3.2_{-0.8}^{+0.5}$ \\
\hline
\end{tabular}
\label{tab:population}
\end{center}
\tablefoot{The available data on the source candidates is collected from \citep{pwn_pop_paper} except for Geminga and B0656+14 \citep{geminga_hawc_paper}. HGPS names are used \citep{HGPS_paper} except for sources located outside of the HGPS; Geminga, B0656$+$14, CTA\,1, 3C\,58, G\,0.9$+$0.1, and the Crab, where instead canonical names are given. The associated pulsar names from ATNF catalogue \citep{Manchester05} are also shown and are numbered in cases where the association is not unique. $\dot E$ and $\tau_{\rm c}$ are the spin down luminosity and characteristic age of the associated pulsar. The sources are sorted by increasing $\dot E$. The estimated distance to the pulsars (which may have large uncertainties) is taken from the ATNF catalogue \citep{Manchester05} unless a different reference is provided. The size is the 68\% containment radius of the TeV emission. $E_{\rm Total} = \dot E \tau_c$, is the estimated total energy in electrons and positrons. $\varepsilon_{\rm e}(\dot E)$ and $\varepsilon_{\rm e}(L_\gamma)$ are the two estimates of energy densities in electrons and positrons (see text in Section \ref{energy_density_estimate}).}

\tablebib{$[1]$ \cite{geminga_dist}; $[2]$ \cite{monogem_dist};  $[3]$ \cite{J1418-609}; $[4]$ \cite{J1849-000};   $[5]$ \cite{J1023-575}; $[6]$ \cite{3C58}}

\end{table*}

\subsection{Estimator using TeV $\gamma$-ray luminosity}
\label{sec:method2}
The second method adopted consists of estimating $\varepsilon_{\rm e}$ from the measured VHE $\gamma$-ray emission of these objects. For Geminga and PSR\,B0656+14, we use the luminosities and spectral indices measured by the HAWC Collaboration over the energy range $E_{\gamma}= 8 - 40$\,TeV~\citep{geminga_hawc_paper}. For all the other PWNe considered here, we use the measurements of their luminosities and spectral indices over the energy range $E_{\gamma}= 1 - 10$\,TeV, as reported in \cite{pwn_pop_paper}. Since the electrons emitting at such high energies only provide a fraction of the total energy density $\varepsilon_{\rm e}$, we need to make a phenomenological assumption on the shape of the electron spectrum outside the energy range where it is constrained by these observations. To do so, we note that the electron spectra inferred for those PWNe that are observed at multiple wavelengths are roughly compatible with broken power-law spectra, with a low-energy break at $E_{\rm low} \sim 100$\,GeV, and a high-energy break at $E_{\rm high} \sim 1 - 10$\,TeV.

At energies $E_{\rm low} \lesssim E_{\rm e} \lesssim E_{\rm high}$, the spectrum is roughly compatible with an $\sim E_{\rm e}^{-2}$ spectrum, and it softens at $E_{\rm e} \gtrsim E_{\rm high}$. At $E_{\rm e} \lesssim E_{\rm low}$, the spectrum is, on the contrary, harder than $E_{\rm e}^{-2}$, implying that the electrons at these low energies contribute little to the total energy density in relativistic electrons --- see for example~\cite{Meyer_100GeV_Break_Crab} for the Crab Nebula. We set $E_{\rm low} = 100$\,GeV, and neglect the contribution of all electrons with $E_{\rm e} < E_{\rm low}$ in our calculations of $\varepsilon_{\rm e}$. In line with the theoretical expectation for particle acceleration at an ultra-relativistic shock with isotropic scattering, we assume that the electron spectrum between $E_{\rm low}$ and $E_{\rm high}$ follows a power-law $\propto E_{\rm e}^{-2.2}$ \citep[e.g.][]{Achterberg2001}. In order to test the dependence of our results on the {\it a priori} unknown value of $E_{\rm high}$, we do our calculations for three different values: $E_{\rm high} = 1$, 3, or 10\,TeV. For each of these values of $E_{\rm high}$, we try two different slopes for the electron spectrum at $E_{\rm e} > E_{\rm high}$. First, we take a spectrum $\propto E_{\rm e}^{-3.2}$, which is consistent with the expectation for a cooled $E_{\rm e}^{-2.2}$ spectrum\footnote{In a steady state, and for electrons continuously injected with a power-law spectrum with index $-p$ and subject to synchrotron (or IC in the Thomson regime) cooling, the resulting population of cooled electrons has a power-law spectrum with index $-p-1$.}. Second, we take a spectrum $\propto E_{\rm e}^{-\Gamma}$, where $\Gamma$ is adjusted individually for each source such that the spectral index of its calculated gamma-ray emission is consistent with the spectral index measured in the relevant energy range.

For each of the six electron spectra described above we calculate for each source the $\gamma$-ray emission from inverse Compton scattering of these electrons on background radiation fields (CMB, UV, optical and IR), using the Equation~(2.48) of~\cite{BlumenthalGould1970} which takes into account Klein-Nishina effects. The spectra of the background UV, optical and IR photons are calculated using the model of~\cite{Tuffs_Rad_Model} for the interstellar radiation fields. These fields are determined individually for each PWN, depending on its location in the Milky Way. Assuming the same source radii $R$ as in the first method, we then infer the electron densities $\varepsilon_{\rm e}$ that are required to fit the observed gamma-ray luminosities in the energy ranges $E_{\gamma} = 1 - 10$\,TeV for the H.E.S.S. paper sources, and $E_{\gamma} = 8 - 40$\,TeV for the two HAWC sources. We note that, since we only calculate typical average energy densities $\varepsilon_{\rm e}$, our calculations are formally equivalent to assuming that the electrons are uniformly distributed within the volumes $V=4\pi R^3/3$ of their sources. When we use the three electron spectra that are $\propto E^{-3.2}$ at $E_{\rm e} > E_{\rm high}$, we only fit the (total) measured gamma-ray luminosity. In contrast, for the three electron spectra with adjustable slopes at $E_{\rm e} > E_{\rm high}$, we simultaneously fit the measured gamma-ray spectral index. For each source, we calculate six values of $\varepsilon_{\rm e}$ --- one per electron spectrum. We provide hereafter the average of these values, and indicate the spread with asymmetric errors.

The results are given in the last column of Table~\ref{tab:population}, and shown in the lower row of Figure~\ref{fig:edensities} as a function of the spin-down power (lower left) or characteristic age (lower right) of the associated pulsar. As in the upper row of Figure \ref{fig:edensities}, the colour of each symbol refers to the source radius $R$ (see the colour bar in the right-hand side), and the area shaded in grey corresponds to the region where $\varepsilon_{\rm e} < 0.1$\,eV/cm$^3$, i.e. where $\varepsilon_{\rm e} \ll \varepsilon_{\rm ISM}$. 
Since the values of $\varepsilon_{\rm e}$ calculated with this second method are derived from measurements, they are likely to provide a better indicator of the actual electron energy densities at the sources than those derived with the estimator of section \ref{sec:method1}. 
Moderately large error-bars are present on the results though, due to the uncertainties on the shape of the electron spectrum. 
One can see that almost all the objects considered here lie above or partly outside the grey band, except for the two HAWC sources (Geminga and PSR\,B0656+14).

\section{Discussion}
\label{sec:discuss}

\subsection{Current halo fraction in TeV-bright PWNe}

Figure~\ref{fig:edensities} demonstrates that the VHE gamma-ray emission from most TeV-bright PWNe is due to electrons that are contained in a region which is influenced energetically or dynamically by the pulsar (i.e. the PWN itself). Therefore, these sources do not constitute proper \lq\lq halos\rq\rq\/, in which the TeV emission arises from a low energy density zone around the PWN. 
It is clear from Figure~\ref{fig:edensities} that the only really unambiguous \lq\lq halo\rq\rq\/ cases in the sample of established TeV-emitting PWN are those powered by Geminga and PSR\,B0656+14, as established by \citet{geminga_hawc_paper}. 
These two HAWC sources stand out as clear halo cases: the electrons responsible for their TeV gamma-ray emission propagate in a region where their contribution to the local total energy density is negligible ($\varepsilon_{\rm e} < 0.1$\,eV/cm$^3$).

We note that there is a strong correlation (coefficient $0.98$) between the two different energy density estimates shown in Figure~\ref{fig:edensities}. This indicates that the gamma-ray properties of PWN are typically more closely related to the current spin-down power, rather than the early lifetime of their pulsars. 
Whilst for the majority of systems the estimator using $\gamma$-ray luminosity results in consistent or lower energy densities than the Case B estimates of section \ref{sec:method1}, for a small number of PWN the energy density is larger, likely due to significant energy injected during the early phase of pulsar evolution.

The threshold in $\varepsilon_{\rm e}$ below which a source can be classified as a halo is, of course, neither sharp, nor universal. 
Several of the sources are close to our indicative grey band. Some of these sources might be reaching the stage of their life when they start to transition towards a halo, although we caution that the presence of incorrect pulsar associations in our sample is not excluded.
Several objects are ambiguous cases in which more careful study of morphology, energetics and environment would be needed for a clear classification. 
The ambiguous cases have characteristic ages consistent with stage~2 of Figure~\ref{fig:pwn_halo_sketch}, and it seems plausible that some of these objects are hybrids, exhibiting halo as well as PWN components. 
HESS\,J1825$-$137 is a candidate for this mixed situation.
The energy-dependence of the morphology strongly suggests advection dominated transport within the PWN, but the fringes of the emission extend to very large distances 
and may indicate the presence of unconfined TeV-emitting particles~\citep{hessj1825_2019}. The emission profile measured with HESS exhibits a change in slope on the Southern side of the nebula, which may hint at a boundary between a high electron energy density core (the PWN) and a surrounding diffusive halo. This effect is more pronounced in linear strips and averages out over azimuthal angle around the pulsar (see Figures\,4 and 5 of \cite{hessj1825_2019}). 
Vela\,X is a well-known example where the TeV IC emission is concentrated within a central region; more extensive than the X-ray synchrotron emission, yet well-contained within the surrounding radio SNR such that we consider this to be a system entering stage~2 \citep{velax_escape,VelaXERN}.

In our sample of TeV-bright systems with firm pulsar associations listed in Table\,\ref{tab:population}, we note that the Crab and Geminga, both frequently used as canonical examples of the class, have the highest and lowest $\dot E$  respectively, suggesting that both are extreme systems rather than typical of the population. 

\subsection{Total halo fraction in Galactic PWNe}

Whilst the fraction of halos in current TeV catalogues is low, these catalogues are clearly biased towards compact, i.e. PWN-like, rather than Halo-like, objects, given the reduction in sensitivity to lower surface brightness emission at fixed flux for resolved objects. In addition, the highest power objects are young (Table~\ref{tab:population}) and would not yet be expected to form halos; only close-by (old, low-power) pulsars are likely to exhibit detectable halos. However, the older systems which do exhibit halos (see e.g. Table~\ref{tab:halos_table} and Figure~\ref{fig:edensities}) are much more numerous, and it is important to consider the entire pulsar population to assess what fraction of the integrated TeV $\gamma$-ray emission is likely to occur within, and outside of, PWN. 

\begin{figure}
\centering
\includegraphics[width=\linewidth]{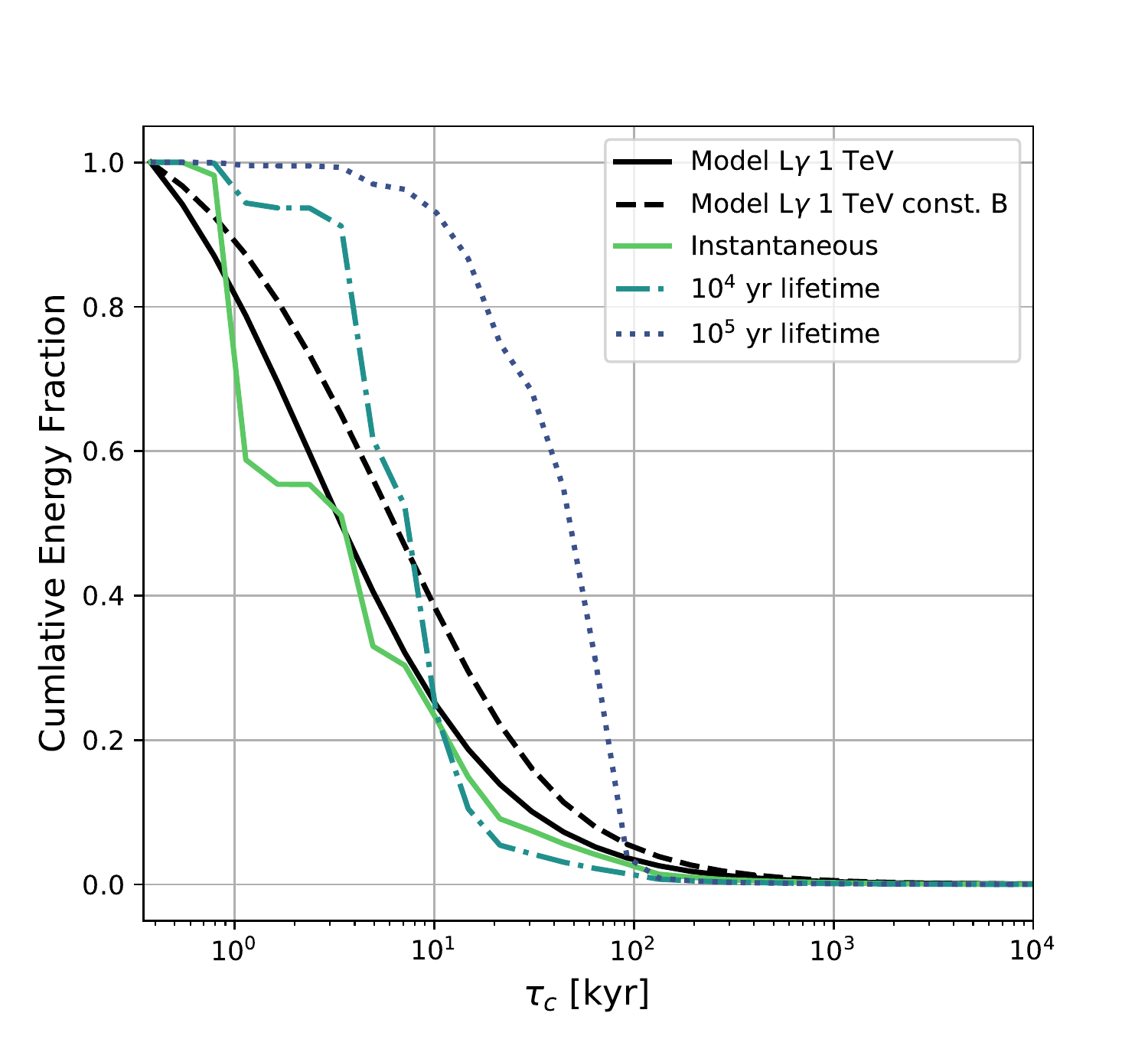}
\caption{Cumulative fraction of the energy output of the known pulsar population (from the ATNF catalogue) created using the instantaneous energy output as well as the output integrated over 10$^4$ and 10$^5$\,year electron lifetimes. For comparison, we show the cumulative energy fraction of the $\gamma$-ray luminosity at 1\,TeV from a generic PWN model as a function of system age, both for a constant magnetic field and a magnetic field evolving in time.}
\label{fig:cumulation_energy_fraction}
\end{figure}

Figure~\ref{fig:cumulation_energy_fraction} shows the cumulative fraction of the energetic output of Galactic ATNF pulsars --- excluding millisecond pulsars\footnote{on the basis that for a spun-up pulsar the characteristic age is not a meaningful indication of the true age.} \citep{Manchester05} as a function of their characteristic age. The energy fraction is extracted from the beaming angle corrected \citep{beamangle}  pulsar $\dot{E}$ shown assuming the current pulsar properties. In order to better assess the contribution of different aged pulsars to the total energy in Galactic electrons the total energy output of the pulsars was integrated over timescales of 10$^4$ and 10$^5$\, years (or the pulsar age if shorter), corresponding to the radiative lifetime in the ISM of around 10 and 1\,TeV electrons respectively. The energy output of the pulsars were assumed to evolve identically as $\left(1 + \frac{t}{\tau_0} \right)^{-\alpha}$, with a $\tau_0$ set to 10$^3$ years and $\alpha = (n+1)/(n-1)$ assumed to be equal to 2 as for a pulsar braking index $n=3$.
Assuming that the power injected in to relativistic electrons evolves in the same manner, a simple evolutionary model using the GAMERA package \citep{gamera} was constructed to predict the expected evolution of the PWN $\gamma$-ray luminosity at 1\,TeV, both in a constant magnetic field environment and for a magnetic field evolving as $(t/\tau_0)^{-0.5}$. The cumulative energy fraction in 1\,TeV $\gamma$-rays from this model is also shown in Figure \ref{fig:cumulation_energy_fraction}. As expected, these curves are approximately consistent with the cumulative energy fraction in the known pulsar population integrated over an electron lifetime of $10^4$ years, corresponding to 10\,TeV electrons. 

Figure~\ref{fig:cumulation_energy_fraction} shows that the contribution of $\gtrsim 10^{5}$ year old pulsars to the total TeV $\gamma$-ray luminosity of PWNe is in general small, and depends on the cooling history of electrons injected early in the life of pulsars, when most of the rotational energy is lost. The $\gamma$-ray luminosity of older halo systems is low, as the majority of the power is injected into the PWN in the early phase of the pulsars evolution, with particle cooling expected prior to escape into the ISM. The overall efficiency for pulsars as cosmic ray electron sources depends on the interplay of the electron lifetime and effectiveness of particle escape.

\subsection{Electron diffusion and escape}

The electron diffusion coefficient measured by HAWC around Geminga and PSR\,B0656+14 is close to the Bohm limit and is two orders of magnitude smaller than the one inferred from the boron-to-carbon ratio \citep{geminga_hawc_paper}. This naturally raises the question of whether the electrons responsible for these two halos probe the \lq\lq unperturbed\rq\rq\/ interstellar turbulence, or, instead, turbulence generated by cosmic-rays escaping from these (or nearby) sources. Cosmic-ray self-confinement around hadronic sources has been studied by, e.g., \cite{Ptuskin2008,Malkov2013,DAngelo2016,Nava2016}. More recently, \cite{evoli_linden_morlino} pointed out that self-confinement could also occur around sources of leptons, suggesting that this may explain the small diffusion coefficient found by HAWC around Geminga and PSR\,B0656+14. 
Our finding in section~\ref{sec:method2} that $\varepsilon_{\rm e} \ll \varepsilon_{\rm ISM}$ around these sources casts doubts on the ability of their electrons to modify substantially the properties of the interstellar turbulence. It cannot be excluded though, that larger electron currents during an earlier phase, or cosmic-rays from a nearby hadronic source, such as the parent SNRs of these pulsars, could have modified the turbulence in the halo regions around these two PWNe. 
However, it is worth noting that there is currently no need to invoke cosmic-ray-driven instabilities to explain the data. In particular, the value of the cosmic-ray diffusion coefficient measured by HAWC is compatible with theoretical expectations for isotropic turbulence with strengths and coherence lengths that are in the relevant ranges for interstellar turbulence \citep{mag_trb_geminga}. If cosmic rays diffuse substantially faster in the Galactic halo than in the local ISM, tensions with the boron-to-carbon ratio might be avoided. In such a scenario, pulsars would remain viable candidates for producing the all-electron spectrum measured at Earth, provided that a nearby undetected pulsar contributes to the high-energy end of the spectrum \citep{undiscovered_pulsar}. In the future, observations of halos around pulsars in different gamma-ray energy ranges should help to constrain the spectrum and nature of the turbulence probed by the emitting electrons. Extended GeV emission has recently been measured with Fermi satellite around Geminga \citep{DiMauro19}.

We discuss here the halo fraction in TeV-bright PWNe, however the formation of halo-like emission at TeV energies due to electron escape is a general phenomenon expected from cosmic accelerators. One possible example is the TeV emission detected beyond the shell in the SNR RX\,J1713.7-3946, hypothesised as due either to a shock precursor or the escape of energetic particles from the shock region \citep{rxj1713}. We note, however, that in this system and other candidate cases for leptonic halos there is often ambiguity with a hadronic emission scenario. 

\subsection{Future prospects}

Future detections of diffusive halos would help considerably to constrain this picture, and the prospects for this were already discussed in~\cite{invisible_pulsars_linden}. New halos may have already been detected: recently, the HAWC Collaboration announced the detection of two more candidate halos - HAWC~J0543+233 around PSR~B0540+23 ($\dot{E}$ = 4.1$\times10^{34}$ erg s$^{-1}$, distance = 1.56 kpc, $\tau_{\rm c}$ = 253 kyr, found in a search for extended sources with disk radius $0.5^\circ$ \citep{atel_j0543}) and HAWC\,J0635+070 around PSR\,J0633+0632 ($\dot{E}$ = 1.2$\times10^{35}$ erg s$^{-1}$, distance = 1.35 kpc, $\tau_{\rm c}$ = 59 kyr, with a Gaussian $1\,\sigma$ extent of $0.65^\circ\pm0.18^\circ$ \citep{atel_j0635}). The pulsars that may be powering these sources are in an advanced state of their evolution, and both fulfill the condition of being older than a few tens of kyr. A rough estimate of their energy densities can be made using the method of section \ref{sec:method1}; yielding values of 0.6 eV/cm$^{3}$ (using the $0.5^\circ$ disk radius in lieu of the 68\% containment radius) and 0.09 eV/cm$^{3}$ (for a 23\,pc 68\% containment radius) for HAWC~J0543+233 and  HAWC~J0635+070 respectively. Whilst these values are certainly consistent with those of Geminga and PSR\,B0656+14 using the estimator based on pulsar properties, it is difficult to provide a more accurate placement in Figure~\ref{fig:edensities} using the method of section \ref{sec:method2} due to the lack of more detailed information about their size and spectral properties. 
Because of the spin-down power and characteristic age of their central source, they should be located inside the grey band corresponding to $\varepsilon_{\rm e} \ll \varepsilon_{\rm ISM}$, and remain clear candidates to be \lq\lq TeV halos\rq\rq, although HAWC\,J0635+070 may be an ambiguous case.  

Where \cite{invisible_pulsars_linden} proposed TeV-bright sources coincident with PWNe as a distinguishable halo phenomenon, most of the sources included in that study are consistent with the classical picture of a middle-age PWN (stage 2 of Figure \ref{sec:PWN_evolution}) with effective confinement, rather than halos. 
From the observational point of view, it must be noted that the X-ray and TeV $\gamma$-ray size of these sources are expected to differ even in fully confined systems, for reasons discussed in section \ref{sec:PWN_evolution}.

\section{Conclusions}

From this study we conclude that the fraction of TeV-bright PWNe exhibiting halo properties is low, however the fraction of PWNe that produce electron halos is presumably large and a generic feature of all pulsars with ages $\gtrsim 10^{5}$ years. In the future, CTA \citep{science_with_CTA}, SGSO/SWGO \citep{SGSO_science_case} and LHAASO \citep{LHAASO_white_paper} will detect new candidates and the number of halos will increase further.  Nevertheless, the total halo fraction of TeV-bright $\gamma$-ray PWN is likely to remain small due to their comparatively low $\gamma$-ray luminosity (Figure\,\ref{fig:cumulation_energy_fraction}) and large angular size, limiting future detections towards local objects. A study of the prospects for future detections of halos by SGSO/SWGO is presented in the Section~4.1 of \cite{SGSO_science_case}.

\bibliographystyle{aa}
\bibliography{bib_file.bib}
\end{document}